\documentclass[twoside,twocolumn,9pt]{article}
\usepackage{extsizes}
\usepackage[super,sort&compress,comma]{natbib} 
\usepackage[version=3]{mhchem}
\usepackage[left=1.5cm, right=1.5cm, top=1.785cm, bottom=2.0cm]{geometry}
\usepackage{balance}
\usepackage{mathptmx}
\usepackage{sectsty}
\usepackage{graphicx} 
\usepackage{lastpage}
\usepackage[format=plain,justification=justified,singlelinecheck=false,font={stretch=1.125,small,sf},labelfont=bf,labelsep=space]{caption}
\usepackage{float}
\usepackage{fancyhdr}
\usepackage{fnpos}
\usepackage[english]{babel}
\addto{\captionsenglish}{%
  
}
\usepackage{array}
\usepackage{droidsans}
\usepackage{charter}
\usepackage[T1]{fontenc}
\usepackage[usenames,dvipsnames]{xcolor}
\usepackage{setspace}
\usepackage[compact]{titlesec}
\usepackage{hyperref}
\usepackage{graphicx}
\usepackage{booktabs}
\usepackage{setspace}

\definecolor{cream}{RGB}{222,217,201}

\begin{document}

\pagestyle{fancy}
\thispagestyle{plain}
\fancypagestyle{plain}{
}

\makeFNbottom
\makeatletter
\renewcommand\LARGE{\@setfontsize\LARGE{15pt}{17}}
\renewcommand\Large{\@setfontsize\Large{12pt}{14}}
\renewcommand\large{\@setfontsize\large{10pt}{12}}
\renewcommand\footnotesize{\@setfontsize\footnotesize{7pt}{10}}
\makeatother

\renewcommand{\thefootnote}{\fnsymbol{footnote}}
\renewcommand\footnoterule{\vspace*{1pt}%
\color{cream}\hrule width 3.5in height 0.4pt \color{black}\vspace*{5pt}} 
\setcounter{secnumdepth}{5}

\makeatletter 
\renewcommand\@biblabel[1]{#1}            
\renewcommand\@makefntext[1]%
{\noindent\makebox[0pt][r]{\@thefnmark\,}#1}
\makeatother 
\renewcommand{\figurename}{\small{Fig.}~}
\sectionfont{\sffamily\Large}
\subsectionfont{\normalsize}
\subsubsectionfont{\bf}
\setstretch{1.125} 
\setlength{\skip\footins}{0.8cm}
\setlength{\footnotesep}{0.25cm}
\setlength{\jot}{10pt}
\titlespacing*{\section}{0pt}{4pt}{4pt}
\titlespacing*{\subsection}{0pt}{15pt}{1pt}

\fancyfoot{}
\fancyfoot[LO,RE]{\vspace{-7.1pt}\includegraphics[height=9pt]{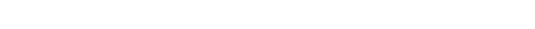}}
\fancyfoot[CO]{\vspace{-7.1pt}\hspace{11.9cm}\includegraphics{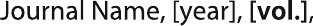}}
\fancyfoot[CE]{\vspace{-7.2pt}\hspace{-13.2cm}\includegraphics{head_foot/RF}}
\fancyfoot[RO]{\footnotesize{\sffamily{1--\pageref{LastPage} ~\textbar  \hspace{2pt}\thepage}}}
\fancyfoot[LE]{\footnotesize{\sffamily{\thepage~\textbar\hspace{4.65cm} 1--\pageref{LastPage}}}}
\fancyhead{}
\renewcommand{\headrulewidth}{0pt} 
\renewcommand{\footrulewidth}{0pt}
\setlength{\arrayrulewidth}{1pt}
\setlength{\columnsep}{6.5mm}
\setlength\bibsep{1pt}

\makeatletter 
\newlength{\figrulesep} 
\setlength{\figrulesep}{0.5\textfloatsep} 

\newcommand{\topfigrule}{\vspace*{-1pt}%
\noindent{\color{cream}\rule[-\figrulesep]{\columnwidth}{1.5pt}} }

\newcommand{\botfigrule}{\vspace*{-2pt}%
\noindent{\color{cream}\rule[\figrulesep]{\columnwidth}{1.5pt}} }

\newcommand{\dblfigrule}{\vspace*{-1pt}%
\noindent{\color{cream}\rule[-\figrulesep]{\textwidth}{1.5pt}} }

\makeatother

\twocolumn[
  \begin{@twocolumnfalse}
\begin{tabular}{m{4.5cm} p{13.5cm} }

& \noindent\LARGE{\textbf{Thin Film Growth Effects on Electrical Conductivity in Entropy Stabilized Oxides }} \\
 \vspace{0.3cm} & \vspace{0.3cm} \\

& \noindent\large{Valerie Jacobson\textit{$^{a,b}$}, David Diercks\textit{$^{b}$}, Bobby To\textit{$^{a}$}, Andriy Zakutayev\textit{$^{a}$}, and Geoff L. Brennecka\textit{$^{b}$}} \\

& \section*{Abstract} \noindent\normalsize{Entropy stabilization has garnered significant attention as a new approach to designing novel materials. Much of the work in this area has focused on bulk ceramic processing, leaving entropy-stabilized thin films relatively underexplored. Following an extensive multi-variable investigation of polycrystalline (Mg$_{0.2}$Co$_{0.2}$Ni$_{0.2}$Cu$_{0.2}$Zn$_{0.2}$)O thin films deposited via pulsed laser deposition (PLD), it is shown here that substrate temperature and deposition pressure have strong and repeatable effects on film texture and lattice parameter. Further analysis shows that films deposited at lower temperatures and under lower oxygen chamber pressure are $\sim$40x more electrically conductive than otherwise identical films grown at higher temperature and pressure. This electronic conductivity is hypothesized to be the result of polaron hopping mediated by transition metal valence changes which compensate for oxygen off-stoichiometry.  \newline

\textbf{Keywords:} Entropy Stabilized, Resistivity, Lattice, SEM, TEM, Morphology

 } 

\end{tabular}

 \end{@twocolumnfalse} \vspace{0.6cm}

 ]

\renewcommand*\rmdefault{bch}\normalfont\upshape
\rmfamily
\section*{}
\vspace{-1cm}


\footnotetext{\textit{$^{a}$~National Renewable Energy Laboratory, 16000 Denver West Pkwy, Golden, Colorado, United States. E-mail: andriy.zakutayev@nrel.gov }}
\footnotetext{\textit{$^{b}$~Colorado School of Mines, 1500 Illinois St, Golden, Colorado, United States. E-mail: geoff.brennecka@mines.edu}}




\section{Introduction}
\begin{figure}[h]
\centering
  \includegraphics[height=4cm]{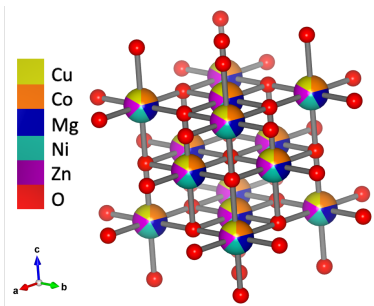}
  \caption{{The films in this study exhibit a rock salt crystal structure in which all five cations are randomly distributed on the cation sublattice with oxygen ions filling the anion sublattice.}}
  \label{fgr:BeachBalls}
\end{figure}

 High entropy alloys are a well established field of work with applications in metallic property optimization\cite{HEA,HEAfracRes,HEAtensile}. In 2015, entropy stabilization was first applied to ceramic oxides\cite{Rost2015}, and a five cation material system, (Mg$_{0.2}$Co$_{0.2}$Ni$_{0.2}$Cu$_{0.2}$Zn$_{0.2}$)O, was found to crystallize reversibly into a uniform rock salt structure when heated to a sufficiently high temperature (>875$^{\circ}$C for the equimolar composition); when quenched to room temperature, the material system retains the rock salt structure. Cation distribution within the entropy stabilized rock salt structure has been shown to be truly random and homogeneous over long range, with some local distortions in the oxygen anion sublattice in order to accommodate the different cation sizes\cite{EXAFS} (Fig.~\ref{fgr:BeachBalls}). This entropy stabilized oxide has since been the focus of many studies, and versions of it have been reported to work well as an Li-ion conductor\cite{IonCon} and anode for Li-ion batteries\cite{LiStorage}, and redox material for thermochemical water splitting\cite{WaterSplit}. Extremely high dielectric constants have also been reported\cite{DConst}. \newline 
  The majority of these studies have been carried out on samples processed as bulk ceramics\cite{bulk1,Rost2015,DConst,JTdistort}; thin film studies are less common. The seminal work on these oxides\cite{RostThesis} showed an inverse relationship between both pO$2$ and substrate temperature on out-of-plane tetragonal lattice parameter in epitaxial thin films of (Mg$_{0.2}$Co$_{0.2}$Ni$_{0.2}$Cu$_{0.2}$Zn$_{0.2}$)O. Another thin film study used pulsed laser deposition (PLD) to show that entropy stabilized thin films of (Mg$_{x}$Co$_{x}$Ni$_{x}$Cu$_{x}$Zn$_{x}$)O can be tailored to engineer long range magnetic order and shows promise for enhancing exchange coupling\cite{ExchCoupling}. PLD has also been used to explore structural stability of the (Mg$_{x}$Co$_{x}$Ni$_{x}$Cu$_{x}$Zn$_{x}$)O system as a function of growth parameters\cite{EpiTFs}. Depositing films using high-energy methods like PLD provides access to much higher effective temperatures (on the order of 30,000 K) than bulk processing allows. Note that the term ``effective temperature'' is used here to describe the estimated temperatures of the plasma generated by laser ablation and not as a measure of cation disorder\cite{Ndione}, as cations are presumably fully disordered in this material system. \newline
 In the current study, PLD is used to fabricate non-epitaxial chemically uniform single phase rock salt thin films of (Mg$_{0.2}$Co$_{0.2}$Ni$_{0.2}$Cu$_{0.2}$Zn$_{0.2}$)O on non-lattice matched substrates. Changing substrate temperature and chamber partial pressure of oxygen enabled tailoring of film texture, lattice parameter, and morphology as well as resulting through-thickness electrical conductivity. Varying deposition temperatures and pressures affects the growth of the deposited thin film, favoring (111) or (002) texture depending on pressure, as well as compressing the lattice constant and changing morphology at lower deposition temperature. Electrical current density measurements through the thicknesses of the films indicate that when (111) crystallographic texturing is coupled with a reduction in lattice constant, the conductivity of the films is significantly higher than under all other conditions.

\section{Experimental Details}
\subsection{Targets}
\begin{figure}[h]
\centering
  \includegraphics[height=7cm]{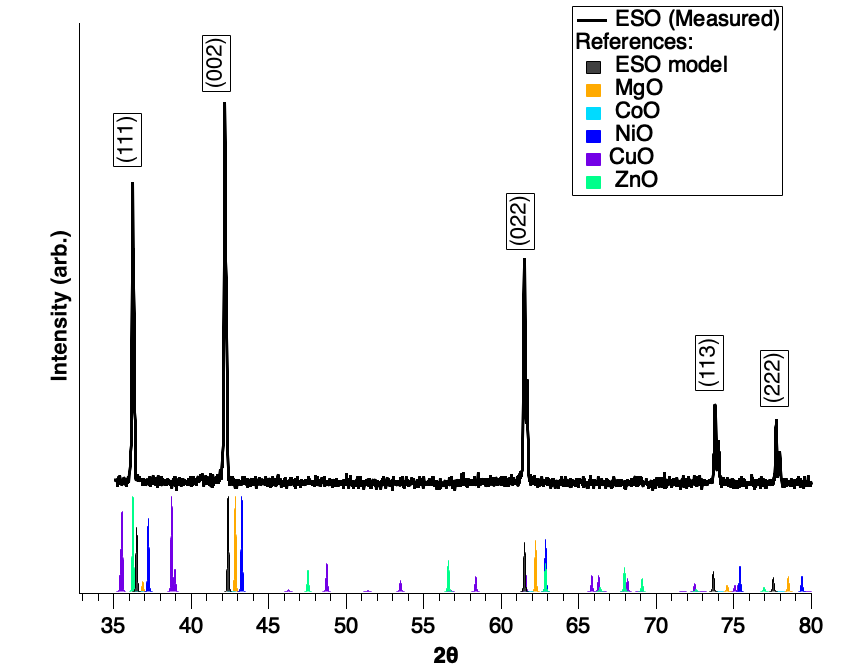}
  \caption{\textbf{}XRD confirms the polycrystalline rock salt structure of the (Mg$_{0.2}$Co$_{0.2}$Ni$_{0.2}$Cu$_{0.2}$Zn$_{0.2}$)O PLD target with a lattice constant of 4.30 \r{A}.\ The five precursor oxide reference patterns\cite{MgO,CoO,NiO,CuO,ZnO} are also shown to confirm that no residual precursor phases are observed.
  \textbf{}}
  \label{fgr:T3XRD}
\end{figure}

Stoichiometric (Mg$_{0.2}$Co$_{0.2}$Ni$_{0.2}$Cu$_{0.2}$Zn$_{0.2}$)O targets were fabricated by mixing equimolar amounts of MgO [Alfa Aesar, 99.99\%], CoO [Alfa Aesar, 99.7\%], NiO [Sigma Aldrich, 99.995\%], CuO [Alfa Aesar, 99.995\%], and ZnO [Alfa Aesar, 99.99\%]. These were roller milled with Y-stabilized ZrO$_2$ milling media for 6 hours. The powder mix was pressed in a 1'' die to 80 MPa using a Carver hydraulic press without binder. Resulting pellets were sintered in air using the following profile: $5^\circ$C/min ramp to $900^\circ$C, hold 6 hrs, 10$^\circ$C/min ramp to $1100^\circ$C, hold 8 hrs. After the 8 hour hold, each pellet was removed from the 1100$^{\circ}$C furnace and air quenched to room temperature. This sintering profile resulted in sufficiently dense ($\approx$87\%) and uniform targets that were confirmed by x-ray diffraction to be polycrystalline single phase rock salt without preferential orientation (Fig. \ref{fgr:T3XRD}). A CuO target was fabricated by pressing and sintering 99.995\% pure powder purchased from Alfa Aesar in a 1'' die to 80 MPa using a Carver hydraulic press without binder. The resulting pellet was sintered in air at $900^\circ$C for 6 hours with a ramp rate of $5^\circ$C per minute. The resulting CuO target was approximately 88\% dense. A PLD target of ZnO (99.9\%) was purchased from Plasma Materials.  
 
\subsection{Pulsed Laser Deposition}
 Single phase films were deposited using a KrF (248nm) excimer laser with pulse rates between 10 Hz and 30 Hz and laser energies ranging from 100 - 350mJ. Film structure was found to be rather insensitive to laser parameters, so a pulse rate of 20 Hz and laser energy of 200mJ were selected as standard for continued study. Both Zn and Cu from the (Mg$_{0.2}$Co$_{0.2}$Ni$_{0.2}$Cu$_{0.2}$Zn$_{0.2}$)O target ablated in slightly lower concentrations than the other cations in the target, so separate CuO and ZnO targets were used to supplement depositions and grow stoichiometric films, with ZnO requiring approximately 1.2\% more laser pulses and CuO requiring approximately 3.5\% more. Substrate temperature and chamber (oxygen) pressure combinations of 200$^\circ$C, 300$^\circ$C, or 450$^\circ$C and 50 mT or 100 mT were explored. The rest of this report focuses on samples deposited with a substrate temperature of $200^\circ$C or $450^\circ$C and oxygen pressure set to 50mT or 100mT; trends reported here are consistent across the rest of the deposition space. Substrates used in this study were Borosilicate [Eagle2000] glass (EXG) and Pilkington NSG TEC 15 Glass with approximately 340nm of fluorinated tin oxide coating (FTO, 13-15 $\Omega$/sq). 

 \subsection{Sample Characterization}
 Samples were characterized using a variety of methods to validate their structure and composition before analyzing properties. Each thin film was measured in 44 different locations in order to confirm consistency of the film across the 2'' x 2'' substrate.

Crystal structure and phase purity of PLD Targets were confirmed using a PANalytical PW3040 X-ray Diffractometer (XRD) and Cu-k$\alpha$ radiation. XRD data for films were collected at the Stanford Linear Accelerator on Beamline 1-5 (measurements calibrated with a LaB$_6$ standard, however BL 1-5 does not correct for tilt of sample. Data presented here is measured to be approximately 0.08 Q low) and with a Bruker D8 Discover with Cu-k$\alpha$ radiation. Composition was confirmed using a Fischer XUV x-ray fluorescence tool (XRF) in addition to energy dispersive spectroscopy (EDS) data collected using a FEI Talos F200X transmission electron microscope (TEM), which was also used for bright field and high angle annular dark field imaging. Top electrical contacts were deposited through a shadow mask using a Temescal FC2000 e-beam tool, with 10nm of Ti and 50nm of Pt. Current density measurements were taken with a Keithley 2400 source meter. Microstructure of the top down was taken with an FEI Nova 630 SEM, while the cross sections were imaged using a Hitachi 4800 SEM in order to manage sample charging. The open-source COMBIgor package for commercial Igor Pro\cite{Combigor} was used for data analysis and visualization.


\section{Results and Discussion}
\subsection{Crystal structure}
Arrays of XRF measurements (Table \ref{tbl:XRF}) confirm that cations are homogeneously distributed in these thin films, at least on a mesoscopic scale. On average, samples are all slightly lower in Cu than other measurable cations; Mg could not be measured using XRF because its atomic mass is too low for the instrument to identify. The slightly lower amounts of Cu do not alter the structure and are consistent across all depositions, so it does not appear to be the cause for the property trends reported here. 
\begin{table}[h]
\small
  \caption{\ Average cation concentration from 44 points across each sample per XRF. }
  \label{tbl:XRF}
  \begin{tabular*}{0.48\textwidth}{@{\extracolsep{\fill}}llllll}
    \hline
    Sample Type & Co & Ni & Cu  & Zn & Mg\\
    (at\%) &&&&&\\
    \hline
    HTLP & 26 $\pm$ 0.4 & 25 $\pm$ 0.4 & 23 $\pm$ 0.4 &25 $\pm$ 0.4 & -- \\
    HTHP & 26 $\pm$ 0.5 & 2 4$\pm$ 0.5 & 23 $\pm$ 0.5 & 26 $\pm$ 0.5 & -- \\
    LTLP & 25 $\pm$ 0.4 & 25 $\pm$ 0.5 & 23 $\pm$ 0.5 & 26 $\pm$ 0.4 & -- \\
    \hline
  \end{tabular*}
\end{table}

\begin{figure}[h]
\centering
  \includegraphics[height=11cm]{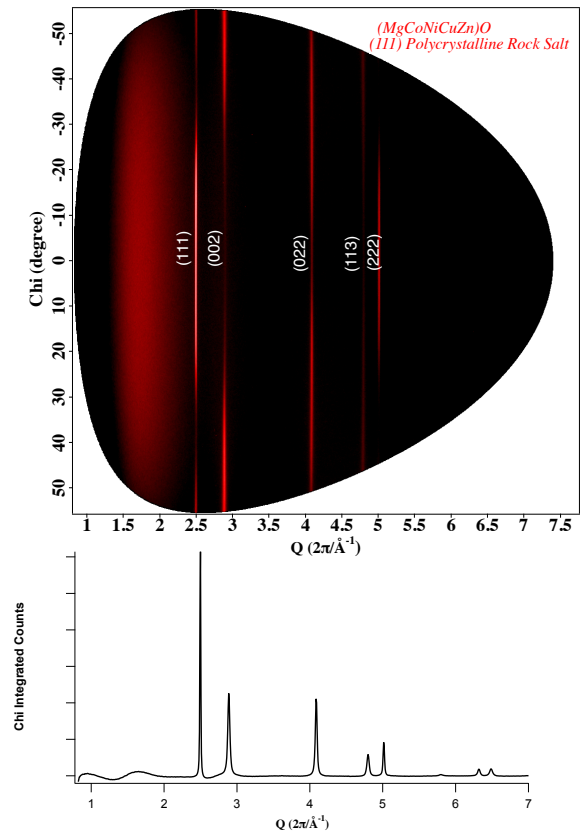}
  \caption{\textbf{Top:}\ High resolution XRD data collected at SLAC on BL 1-5 show the (111)-textured polycrystalline growth of a cubic rock salt with moderate (111) crystallographic texture on a borosilicate glass substrate. \textbf{Bottom:}\ Integrated counts across approximately 4 degrees through the center of the measured Chi emphasize the (111) texture of the sample.}
  \label{fgr:SLACpolyxtal}
\end{figure}

\begin{figure}[h]
\centering
  \includegraphics[height=12cm]{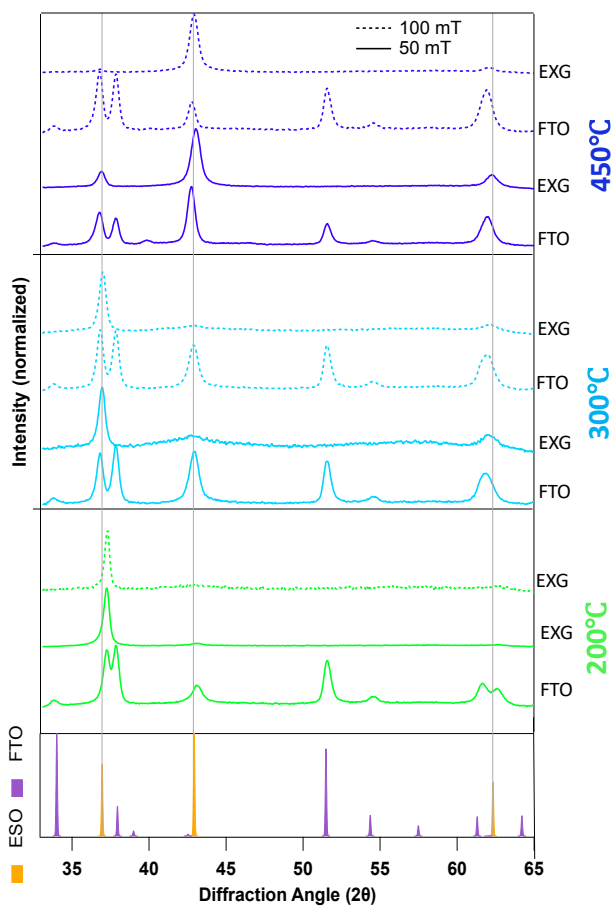}
  \caption{\textbf{}Crystallographic texturing was found to change as a result of deposition temperature and partial pressure of oxygen. Overall, samples grown at 450$^\circ$C have a tendency to grow with an (002) texture when grown on either EXG or FTO substrates and at both 50 mT and 100 mT pO$_2$. 
  Samples grown at 300$^\circ$C show (111) texturing at both 50 mT and 100 mT and on either EXG or FTO substrates. Samples grown at 200$^\circ$C with 50 mT or 100 mT all show (111) texturing along with peaks at higher $2\theta$, indicating a shorter lattice constant for these films. Samples grown on EXG glass versus FTO-coated glass are designated by ``EXG'' and ``FTO''.}
  \label{fgr:PvTXRD}
\end{figure}

XRD confirms that the rock salt structure of the films matches that of the desired rock salt ESO (Mg$_{0.2}$Co$_{0.2}$Ni$_{0.2}$Cu$_{0.2}$Zn$_{0.2}$)O (Fig. \ref{fgr:SLACpolyxtal}), similar to the target (Fig. \ref{fgr:T3XRD}), and suggests that all constituent cations are most likely randomly and homogeneously dispersed throughout the the cation FCC sublattice.\cite{EXAFS,RostThesis}. 
Films grown on EXG at $200^\circ$C under 50mT pO$_2$ exhibit moderate levels of texturing in the (111) growth direction (Fig. \ref{fgr:SLACpolyxtal}) where the corrected cubic lattice constant \emph{a} is 4.22 \r{A}\.\ It is interesting to note the difference in the thin film lattice constant with that of the (Mg$_{0.2}$Co$_{0.2}$Ni$_{0.2}$Cu$_{0.2}$Zn$_{0.2}$)O target (4.30 \r{A}), indicating that the thin film growth on an amorphous substrate exhibits some lattice compression over that of the target material in all directions. 
Figure~\ref{fgr:PvTXRD} shows changes in film texture with substrate temperature and chamber pressure of oxygen. Note that all of the lattice constants for samples shown here are smaller than that of the bulk ESO target lattice constant (4.3 \r{A}, Fig. \ref{fgr:T3XRD}). The samples in this study do not show the specific out of plane lattice compression identified in earlier work on this material system\cite{RostThesis} because the films here are not epitaxially constrained by the substrates; rather they show a cubic compression where both the \emph{a} and \emph{c} lattice constants are slightly compressed relative to the bulk ceramic.  \newline

\begin{figure}[ht]
\centering
  \includegraphics[height=5.3cm]{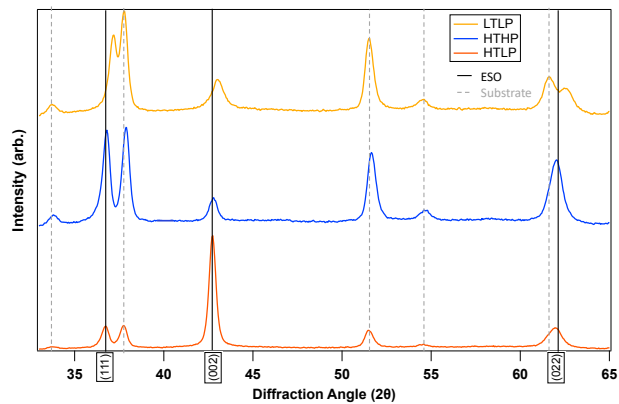}
  \caption{Two dimensional XRD data collected using a Bruker D8 Discover shows the lattice shift in the ESO rock salt peaks relative to the FTO peaks. ESO peaks are identified by solid grey lines and FTO peaks by dotted grey lines. The LTLP samples show a shift to higher 2$\theta$ for ESO peaks, verifying a smaller lattice constant.}
  \label{fgr:3types}
\end{figure}

Figure~\ref{fgr:3types} illustrates different types of growth observed in this study, and their respective definitions. 
Three distinct growth types have been identified based on texture and lattice constant and will be the focus for the remainder of this paper.  Each of these three types of films is grown on FTO substrates and will be referred to as ``HTHP''(Higher Temperature, Higher Pressure), ``LTLP'' (Lower Temperature, Lower Pressure), and ``HTLP''(Higher Temperature, Lower Pressure) and are defined as follows: HTHP and HTLP both have a lattice constant of 4.22 \AA\, while LTLP samples have a lattice constant of 4.19 \r{A}. The Lotgering Factor\cite{LF,JonesLF} is used here to quantify the degree of crystallographic texturing; LF = 0 is purely random, LF = 100\% is perfectly textured. All LF values in this paper were calculated from an integration across 4 degrees of Chi from 2d detectors. Films designated as ``HTHP'' show a (111) texture with an LF = 31\% and are grown at 100mT pO$_2$ and 450$^\circ$C; ``LTLP'' films show (111) texturing as well, but have an LF = 41\% and are grown at 50mT pO$_2$ and 200$^\circ$C; ``HTLP'' films show (002) texture with an LF = 36\% and are grown at 50mT pO$_2$ and 450$^\circ$C. 
This information is summarized in Table \ref{tbl:Summary}.\newline
The definition for lattice strain\cite{LNOeprops} is \( \frac{(c-c_{bulk})}{c_{bulk}} \), where $c$ is the out-of-plane lattice constant and $c_{bulk}$ is the bulk lattice constant associated with strain-free polycrystalline growth (here we use the lattice constant from the ESO target of 4.30 \r{A} as our bulk value). We can apply this model to ESO data as a way to quantify the degree of compression happening in these films relative to bulk samples. The comparative lattice constant for individual thin film samples is calculated for a cubic crystal lattice and derived from individual XRD scans using the 2$\theta$ positions of the center of at least two peaks from different crystal plane families. This lattice compression for LTLP samples indicates some fundamental difference in these samples from the HTHP and HTLP samples.  All LTLP samples grown on FTO have an isotropic compression greater than $2.7\%$, while HTLP and HTHP samples have an isotropic compression no greater than $2.1\%$.

\subsection{Microstructure}
\begin{figure}[h]
 \centering
  \includegraphics[height=5.2cm]{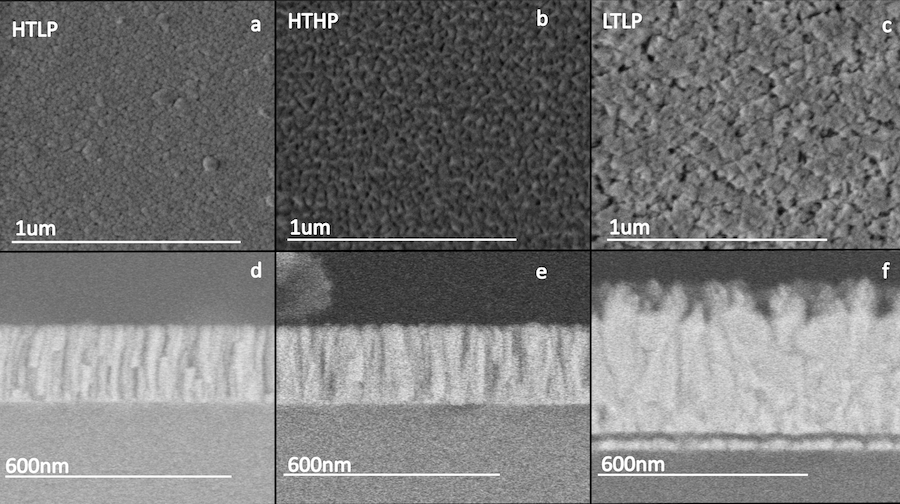}
  \caption{\textbf{} \textbf{a-c)}SEM shows different surface morphologies for each of the three crystallographic orientations.
    \textbf{d-f)} Cross-sectional SEM shows that the HTLP and HTHP samples have a fairly columnar grain structure, while the LTLP textured samples have overlapping grains and less uniformity across the film thickness. *Note: LTLP films were grown with 30\% more laser pulses, making them thicker, as seen above.}
  \label{fgr:SEM3t}
\end{figure}

SEM images revealed notably different microstructures among these three growth types. The HTLP textured samples, grown at $450^\circ$C and 50mT O$_2$, have almost circular grains growing in a largely columnar fashion, as seen in Fig.~\ref{fgr:SEM3t}a,d. HTHP samples, grown at $450^\circ$C and 100mT O$_2$, also show columnar grains, however these grains exhibit a triangular shape at the surface consistent with the cube corners of a (111)-textured rock salt, shown in Fig.~\ref{fgr:SEM3t}b,e. LTLP samples, grown at 200$^\circ$C and 50mT O$_2$, do not exhibit columnar growth, and grains show no consistent morphology, seen in Fig.~\ref{fgr:SEM3t}c,f. It is not unusual for films grown at higher temperatures to exhibit more regular columnar growth as the higher substrate temperature provides additional mobility for atoms to settle into a somewhat lower energy configuration before being fully ``quenched''. The HTHP films are grown under a higher partial pressure of oxygen and higher substrate temperature. Higher pressure leads to more oxidizing conditions during growth, while the higher temperature leads to more reducing conditions according to ideal gas law, but also may give the material the kinetic energy needed to incorporate oxygen into the lattice. The kinetic limitations of the lower substrate temperature coupled with a lower growth pO$_2$ suggest that oxygen incorporation into the lattice may be different in such samples, leading to different oxygen off-stoichiometry.

\begin{figure}[h]
\centering
  \includegraphics[height=5cm]{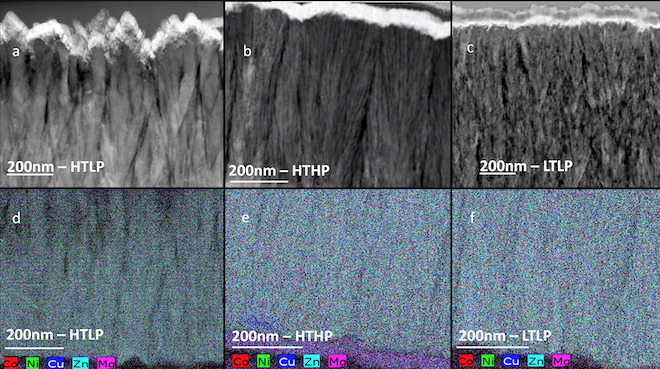}
  \caption{\textbf{}Film microstructure as shown in HAADF images is consistent with columnar grains seen in SEM.  \textbf{a)}  Samples with an HTLP texture show a higher degree of surface roughness in addition to a lower degree of collimation than in the HTHP samples but a higher degree of collimation than the LTLP samples. \textbf{b)} HTHP samples show a high degree of collimation in grain structure,  \textbf{c)} LTLP samples show grains to be less ordered than in the HTHP and HTLP samples. \textbf{d-f)} Cation distributions for all three types of crystallographic texturing are uniform and homogeneous as shown by EDS mapping.}
  \label{fgr:TEMsp}
\end{figure}

\begin{figure}[h]
\centering
  \includegraphics[height=5cm]{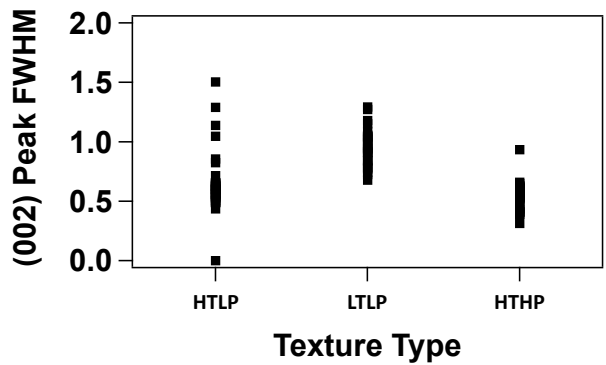}
  \caption{\textbf{}None of the three sample types show a shift in XRD data FWHM calculations, indicating that cation segregation at higher temperatures is unlikely to be the underlying cause of the observed differences in resistance.}
  \label{fgr:noPhaseDecomp}
\end{figure}

Uniformity of the cation distributions is shown with an EDS map across a HAADF image in Fig. \ref{fgr:TEMsp}. The microstructures and z-contrast seen in Fig. \ref{fgr:TEMsp} are consistent throughout the thickness of each film, suggesting that any variation in electrical response between these three types of samples arises from something other than cation segregation. A TEM and atom probe tomography study on bulk samples of the same composition\cite{Diercks} showed that annealing could lead to Cu segregation, but no evidence of such segregation is observed here. \newline
It has been established\cite{RostThesis} that phase decomposition of bulk samples begins in an annealing process around $600^\circ$C, which may suggest some likelihood of phase segregation for film samples grown at higher temperatures. Such segregation is accompanied by XRD peak broadening\cite{JTdistort}. Films in this study exhibit consistent full width half maximum values for XRD peaks across all deposition temperatures (Fig. \ref{fgr:noPhaseDecomp}). Thus, neither XRD nor TEM (imaging and EDS) show any evidence of cation segregation in these films.   \newline 

\subsection{Electrical Properties}

\begin{figure}[]
\centering
  \includegraphics[height=6cm]{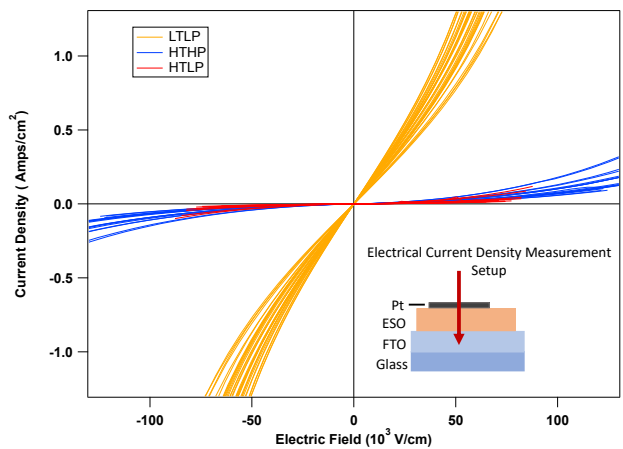}
  \caption{The resistance of HTLP and HTHP samples is much higher than in the LTLP textured samples. Multiple lines for each sample type are the result of each of the 44 points mapped across each film. Reported values are an average of at least 30 measurements for each sample type. Inset: Schematic of measurement configuration. }
  \label{fgr:IVdata}
\end{figure}

Films of approximately 1 $\mu$m thickness were grown on FTO coated borosilicate glass in order to enable through-thickness current density measurements. Measuring the conductivity of these films as a function of applied voltage reveals different electrical behavior for the LTLP samples from the other two types of films. Measurements collected on HTLP samples reveal a nonlinear resistance on the order of 3.42$\pm$0.175 M$\Omega$cm. HTHP samples show a nonlinear resistance on the order of 1.2$\pm$0.05 M$\Omega$cm. Samples with the LTLP textured structure are much more conductive, with an average measured resistance of 51$\pm$0.65 k$\Omega$cm as seen in Fig. \ref{fgr:IVdata}. The nonlinearity was initially hypothesized to be a result of self heating of the sample during the measurement. After a series of repeated current density measurements over a range of different frequencies and duration, this appears unlikely to be the cause because the measurements proved consistent across all acquisition timing parameters. \newline 
Samples deposited at the lowest substrate temperature ($200^\circ$C) and lowest oxygen pressure (50 mT) were the most electrically conductive. Because these films were grown at the lowest substrate temperature, they are the most kinetically limited sample set. Because of the kinetic limitations, it is likely that these films grew with an oxygen off-stoichiometry, which could be compensated by a change in average cation valence. In this case, the most likely mechanism of electrical conductivity is polaron hopping; the hopping distance is shorter for polarons in the compressed lattice, making the hop more likely to occur and increasing electrical conductivity. There is also a clear correlation between higher electrical conductivity (Fig. \ref{fgr:IVdata}) and non-columnar microstructure (Fig. \ref{fgr:SEM3t}), but the exact origin of this correlation is not clear.

The results presented here are consistent with previously published data as represented by Table \ref{tbl:Summary}. One publication has reported electrical conductivity\cite{DConst} in bulk samples to be 5 M$\Omega$cm, which is consistent with the resistive measurements in the present study. Presumably, the bulk samples from this previous publication\cite{DConst} also have a random microstructure similar to that of the LTLP samples, but the improved density of a thin film over a bulk processed sample would increase electrical conductivity and explain the lower resistance values measured in our thin film samples. Other publications do not present resistivity data, however they do show a consistent lattice constant across non-epitaxial samples. The epitaxial samples grown on MgO$_{(001)}$ substrates are reported to have a much smaller lattice constant (4.15 \AA\ ). Nanopowder samples show some tetragonal distortion not seen in other bulk studies.\newline

\section{Conclusions}
Single phase rock salt (Mg$_{0.2}$Co$_{0.2}$Ni$_{0.2}$Cu$_{0.2}$Zn$_{0.2}$)O samples can be deposited efficiently using PLD, and temperature and partial pressure of oxygen can be used to control crystallographic texturing. Thin film samples grown on FTO with a lattice constant of 4.22 ~\AA\ and either (111) or (002) texturing (HTHP or HTLP, respectively) demonstrate a nonlinear electrical resistance between 1 and 3 M$\Omega$cm, while (111) textured films with a lattice constant of 4.19 ~\AA\ (LTLP) are significantly more conductive, with resistivity values on the order of 50 k$\Omega$cm. The primary difference in microstructure for resistive versus conductive films is that the conductive films have a more random grain structure with nucleation occurring throughout the films, and the resistive films show columnar grains with nucleation occurring primarily at the substrate. It is unclear whether this is correlation or causation. The smaller lattice constant also indicates a shorter hopping distance for any induced charge in the material system. This, coupled with a likely oxygen off-stoichiometry during growth, makes polaron hopping a likely candidate for the conduction mechanism. \newline
Ultimately, this work indicates that thin film samples grown on FTO at 50mT pO$_2$ and a substrate temperature of 200$^\circ$C are more electrically conductive than samples grown at higher substrate temperatures with a more organized microstructure by approximately 40x. The data presented here are consistent with previous reports on this material, but also shows that manipulating microstructure and crystal lattice can have dramatic effects on electrical conductivity.

\begin{table*}[t]
\small
  \caption{\ Summary of growth conditions and characteristics for all samples grown on FTO substrates and compared to previous works. *Note: LF and microstructure information for previous works have been approximated based on data provided in articles.}
  \label{tbl:Summary}
  \begin{tabular*}{\textwidth}{@{\extracolsep{\fill}}lllllllll}
    \hline
    Sample & Pressure      & Temperature & Crystal & Lattice        & Lotgering & Average Lattice & Microstructure & Resistance \\
    Type   & (mT P${O_2}$) & ($^\circ$C) & Texture & Constant (\AA) & Factor    & Compression(\%) &                &(k${\Omega}$cm) \\ 
    \hline 
    HTLP & 100 & 450 & (002) & 4.22 & 30\% & 1.81 $\pm$ 0.24 & Columnar & 3400 $\pm$ 175  \\
    HTHP & 50 & 450 & (111) & 4.22 & 31\% & 1.59 $\pm$ 0.13 & Columnar & 1200 $\pm$ 50 \\
    LTLP & 50 & 200 & (111)& 4.19 &  41\% & 2.75 $\pm$ 0.06 & Random & 51 $\pm$ 0.65 \\
    Bulk Polyxtal\cite{DConst} & 20\% & 1000 & N/A & 4.22 & 0\% & 1.86 & Random & 5000  \\
    Nanoparticles\cite{LiStorage} & 20\% & 1000 & N/A & a=4.17; c=4.2 &  0\% & 2.33 & Random & --\\
    MgO(001) epi\cite{ExchCoupling} & 50 & 300 & (002) & 4.15 & 100\% & 3.49 & Epitaxial & --  \\
    TF on Al$_2$O$_3$\cite{RostThesis} & 50 & 500 & N/A & 4.25 &  0\% & 1.16 & Random & --  \\
    
    
  \end{tabular*}
\end{table*}

\section*{Conflicts of interest}
There are no conflicts to declare.

\section*{Acknowledgements}
This work was authored in part by the National Renewable Energy Laboratory (NREL), operated by Alliance for Sustainable Energy LLC, for the U.S. Department of Energy (DOE) under contract no. DE-AC36-08GO28308. Work at NREL by A.Z. was funded provided by the Office of Energy Efficiency and Renewable Energy (EERE), under Fuel Cell Technologies Office (FCTO), as a part of HydroGEN Energy Materials Network (EMN) consortium. Work at CSM by G.B. and V.J. was partially supported by the National Science Foundation (DMR-1555015 and DMREF-1534503). The use of Stanford's Synchrotron Radiation Lightsource, SLAC National Accelerator Laboratory, was supported by DOE‐SC‐BES under Contract No. DE‐AC02‐76SF00515. We would like to thank K. Talley and A. Mis for assistance in data processing with COMBIgor;  P. Walker for helping to streamline the FIB liftout process; K. Gann for technical support; A. Mehta and M.S. Perera for  support on SLAC's BL 1-5; B. Gorman for TEM instruction. The views expressed in the article do not necessarily represent the views of the DOE or the U.S. Government.



\balance


\bibliography{rsc} 
\bibliographystyle{rsc} 

\end{document}